\documentstyle[preprint,aps]{revtex}
\begin{document}
\draft

\title{Sharp and Smooth Boundaries of Quantum Hall Liquids}
\author{C. de C. Chamon and X. G. Wen}
\address{Department of Physics, Massachusetts
Institute of Technology, \\ Cambridge, Massachusetts 02139}
\maketitle
\begin{abstract}
We study the transition between sharp and smooth density distributions
at the edges of Quantum Hall Liquids in the presence of interactions.
We find that, for strong confining potentials, the edge of a $\nu=1$
liquid is described by the $Z_F=1$ Fermi Liquid theory, even in the
presence of interactions, a consequence of the chiral nature of the
system.  When the edge confining potential is decreased beyond a
point, the edge undergoes a reconstruction and electrons start to
deposit a distance $\sim 2$ magnetic lengths away from the initial QH
Liquid.  Within the Hartree-Fock approximation, a new pair of branches
of gapless edge excitations is generated after the transition. We show
that the transition is controlled by the balance between a long-ranged
repulsive Hartree term and a short-ranged attractive exchange term.
Such transition also occurs for Quantum Dots in the Quantum Hall
Regime, and should be observable in resonant tunneling experiments.
Electron tunneling into the reconstructed edge is also discussed.
\end{abstract}

\section{Introduction}

In Fractional Quantum Hall (FQH) states there are no bulk gapless
excitations; the only gapless modes are edge states, which are
responsible for non-trivial transport properties at low temperatures.
Edge states arise naturally in real samples, as the two dimensional
electron gas (2DEG) is confined in a finite region. The manner in
which the 2DEG is confined determines the structure of the electronic
density on the borders of the sample, and rich structures may appear.
For smooth edges, strips of compressible and incompressible FQH states
may be formed \cite{Kouwenhoven,Chang}.  The structure of these smooth
edges have been studied mainly by focusing on the electronic density
distribution at large length scales, where it is reasonable to use a
semi-classical approach \cite{Chang,Beenakker,Chklovskii}. This
approach consists in assuming that the electron density $n$ varies
slowly enough, so that one can use bulk values for the internal energy
$u(n)$, which has cusps for $n$ corresponding to fractional filling
factors.  An improved calculation using Hartree-Fock approximation was
done in Ref. \cite{GH} at finite temperature, which agrees very well
with the electrostatic calculation \cite{Chklovskii}.

In the opposite extreme, the electron density varies sharply at the
edge, and there is no room for the formation of incompressible strips.
The sharp edge of the $\nu=1$ state is described by a 1D (chiral)
Fermi Liquid, in which the occupation in the momentum space has a
sharp drop at the Fermi momentum. This sharp drop in the momentum
space is related to the fast drop of the electron density near the
edge.

One naturally questions how the sharp edge picture evolves into the
smooth edge picture as the edge potential becomes smoother. As an
interacting 1D system, the Fermi edge of the $\nu=1$ state may have
the following possible singularities displayed in Figure 1, such as a
Fermi Liquid singularity, with a $Z_F=1$ discontinuity (Fig. 1a) or a
renormalized $Z_F<1$ discontinuity (Fig. 1b), or a Luttinger Liquid
singularity (Fig. 1c).  One scenario is that the sharp edge and the
smooth edge are connected by the distributions in Fig. 1b or Fig. 1c.
As the edge potential becomes smoother, the occupation distributions
in Fig. 1b and Fig. 1c also get smoother. According to this picture,
the smooth edge of the $\nu=1$ state contains one branch of gapless
edge excitations which is described by a renormalized Fermi Liquid.
The smoother the edge, the stronger the renormalization. However, the
calculations presented in this paper suggest a new scenario for some
natural confining potentials.  We find that the chiral nature of this
one dimensional system play an important role in determining the form
of the singularity.  Due to the chirality, the $Z_F=1$ edge is very
stable.  Fig. 1a correctly describes the edge structure for a range of
edge potentials even for interacting electrons. However, as the edge
potential is smoothened beyond a certain point, the edge undergoes a
reconstruction. A pair of edge branches moving in opposite directions
is generated, and the occupation distribution in Fig. 1a changes into
the one in Fig. 2, which contains three Fermi points.  The occupation
$\langle n_k\rangle$ has algebraic singularities at these Fermi
points. This new scenario has received some support from exact
calculations on small systems.

In this paper we focus on the structure of the electronic occupation
density distribution at the boundary of a $\nu=1$ liquid in the
presence of interactions. We work with a Hilbert space restricted to
the first Landau level, and in the main part of the paper we assume
that the spins are fully polarized.  The droplet is confined by its
interaction with an underlying positive background (one way to
introduce a confining potential).  The paper is organized as follows.
In section II we introduce the 1D interacting version for the problem,
and discuss the importance of chirality in determining the structure
of the edge singularity. In section III we present exact numerical
results for small systems, which support the picture that, before a
discontinuous transition occurs, the chiral edge system is reasonably
well described within the Hartree-Fock approximation. In section IV we
study the effects of this transition for Quantum Dots, including
possible experimentally observable effects.  Finally, in section V we
investigate the consequences of the transition on the dynamics of edge
excitations, and in section VI the consequences to electron tunneling
into the reconstructed edges.

\section{The 1D Interacting Model and Consequences of Chirality}

A system of interacting particles in a 2-D QH droplet can be mapped
into a one dimensional problem by enumerating the single particle
wavefunctions of the first Landau level. The Hamiltonian of the
interacting theory is
\begin{equation}
H=\sum_{\lambda,\lambda'}\epsilon_{\lambda,\lambda'} \
c^{\dagger}_{\lambda} c^{}_{\lambda'} +
\sum_{\lambda_1,\lambda_2,\lambda_3,\lambda_4}
V_{\lambda_1,\lambda_2,\lambda_3,\lambda_4}\ c^{\dagger}_{\lambda_1}
c^{}_{\lambda_2} c^{\dagger}_{\lambda_3} c^{}_{\lambda_4}\ \ ,
\end{equation}
where
\begin{eqnarray}
\epsilon_{\lambda,\lambda'} &=&\int dx^{2}_{1} dx^{2}_{2}
\rho(\vec{x}_1)V(|\vec{x}_1-\vec{x}_2|)\phi^{*}_{\lambda}(\vec{x}_2)
\phi_{\lambda'}(\vec{x}_2) \nonumber \\
V_{\lambda_1,\lambda_2,\lambda_3,\lambda_4} &=&\frac{1}{2}\int
dx^{2}_{1} dx^{2}_{2} \phi^{*}_{\lambda_1}(\vec{x}_1)
\phi_{\lambda_2}(\vec{x}_1)
V(|\vec{x}_1-\vec{x}_2|)\phi^{*}_{\lambda_3}(\vec{x}_2)
\phi_{\lambda_4}(\vec{x}_2)\ \ . \nonumber
\end{eqnarray}
The dispersion $\epsilon_{\lambda,\lambda'}$ is determined by a
background charge $\rho(\vec{x})$, which we use to control the
confining potential. The $\phi_{\lambda}$'s are the single particle
wavefunctions, labeled by the quantum number $\lambda$. For example,
in the symmetric gauge, $\lambda$ stands for the angular momentum
quantum number $m$, with $\phi_m(x,y)=\frac{1}{\sqrt{\pi}}\frac{z^m}{
\sqrt{m!}}e^{-|z|^2/2}$ and $z=\frac{x+iy}{\sqrt{2}}$ (throughout the
paper we work in units of magnetic length $l_B=1$). The wavepacket
$\phi_m$ is centered in a circle of radius $R=\sqrt{2m}$. In the
Landau gauge, $\lambda$ denotes the linear momentum in the $x$
direction $k_x$, with
$\phi_{k_x}(x,y)=\frac{1}{(\sqrt{\pi}L)^{1/2}}e^{ik_{x}x}e^{-(y-k_x)^2/2}$,
and the wavepacket $\phi_{k_x}$ is centered at $y=k_x$ ($L$ is the
size of a system subject to periodic boundary conditions).

Let us consider for now backgrounds $\rho(\vec{x})$ that are invariant
under certain symmetry transformations, such as rotations (if we are
studying a circular droplet, using the symmetric gauge) or
translations along the $x$ direction (if we are studying a long strip,
using the Landau gauge). In this case we have
$\epsilon_{\lambda,\lambda'}=\epsilon_{\lambda}\
\delta_{\lambda,\lambda'}$.  Impurities break such symmetries, and
their effect will be considered later in the paper.  Because the
interaction $V(|\vec{x}_1-\vec{x}_2|)$ depends only on the distance
between $\vec{x}_1$ and $\vec{x}_1$, it is also invariant under these
symmetries, and thus we can rewrite the Hamiltonian as
\begin{equation}
H=\sum_{\lambda}\epsilon_\lambda c^{\dagger}_{\lambda} c^{}_\lambda +
\sum_{\delta,\lambda,\lambda'} V(\delta,\lambda,\lambda')\
c^{\dagger}_{\lambda+\delta} c^{}_{\lambda} c^{\dagger}_{\lambda'}
c^{}_{\lambda'+\delta}\ \ .\label{h1d}
\end{equation}
It is this 1D interacting model that will be the basis of our study of the
$\nu=1$ droplet. The question we want to address is how to determine the ground
state occupation number $\langle c^{\dagger}_{\lambda} c_\lambda \rangle$ for
this theory.

Without loss of generality, let us focus now on the problem of a strip
with length $L$ and periodic boundary conditions (equivalently, a
cylinder of circumference $L$), using the Landau gauge. The QH Fluid
lies on the surface of the cylinder, between its left ($L$) and right
($R$) boundaries (see Fig. 3).

In a typical 1D interacting theory we have non-Fermi Liquid behavior;
the Fermi discontinuity is destroyed by the interactions, and the
system is better described as a Luttinger Liquid. Notice, however,
that the Hamiltonian in Eq. (\ref{h1d}) has a peculiar difference from
the usual 1D Hamiltonian of an interacting system in the sense that
the effective scattering potential $V$ depends not only in the
momentum transfered $\delta$, but also in the momentum configuration
({\it i.e.}, $\lambda$ and $\lambda'$) of the scattered particles. The
Luttinger Liquid behavior is caused by the coupling between
particle-hole excitations in the two distinct Fermi points. Now, for
our system described in Eq. (\ref{h1d}), the two Fermi points
correspond to $\lambda_L$ and $\lambda_R$, at the two boundaries of
our droplet. These two points are spatially separated, and the matrix
elements for coupled particle-hole excitations near these points
should go to zero as the distance between the boundaries is increased.
In the limit of infinite separation, the two edges are decoupled, and
we can describe the system as containing two different types of
fermions, $R$ and $L$, with one Fermi point each. More precisely, we
can describe the particles as being in a Dirac sea that is filled as
we move inwards to the bulk. Indeed, in topologies such as a simply
connected droplet, like a disk, we only have one boundary, and the
Dirac sea description is exact. Such theories fall within what we call
``Chiral Luttinger Liquids''.

The 1D chiral theory has a special property that the occupation
distribution of the ground state can have a Fermi discontinuity (Fermi
Liquid) {\it even in the presence of interactions}.  In fact, for
certain values of interaction strength and single particle dispersion,
the ground state may have a perfect Fermi distribution with $Z_F=1$!
This is because the momentum occupation with $Z_F=1$ is always an
eigenstate of the {\it interacting} Hamiltonian (\ref{h1d}) in the
limit of infinitely separated edges. This is easier seen in the filled
Dirac sea description of the $L$ and $R$ fermions. Take, for example,
the $R$ branch, for which the unique minimum total momentum eigenstate
is the one that has all single particle levels to the left of the edge
occupied. Because total momentum commutes with the Hamiltonian, and
this state is the only one with minimum total momentum, it must also
be an eigenstate of energy, possibly the ground state for some edge
potential.  Notice that the occupation distribution of this state
corresponds exactly to a $Z_F=1$ Fermi Liquid occupation.  One should
contrast this case with a non-chiral 1D system, where clearly the
Fermi gas distribution is an eigenstate of zero total momentum, but it
is not the only one, and thus not necessarily an eigenstate of energy.
Again, chirality plays a key role.

The next step is to understand how the occupation distribution evolves
as we smoothen the confining potential. One way to assemble a sharp
distribution is by simply laying the electron gas on top of a
similarly sharp positively charged background, and one way to try to
destroy this sharp distribution is to smoothen the positive
background.  Notice that the perturbation we include by changing the
background is not in the form of an additional interaction between the
particles, but of a change in the one particle dispersion
$\epsilon_\lambda$.  We will show that the occupation distribution has
the tendency to remain sharp, due to a balance between a repulsive
long range Hartree term and an attractive short range Fock term, and
also due to the special stability of the $Z_F=1$ chiral Fermi Liquid.
The sharp distribution eventually becomes unstable, and the Fermi
surface is destroyed, as a lump of particles detach and form two more
edges which destroy the chirality.  We will show that the Hartree-Fock
approximation seems to contain the relevant ingredients to describe
this transition.

\section{Exact Results for Small Systems}

In the Landau gauge, the dispersion due to the background charge, and
the matrix elements $V$ in Eq. (\ref{h1d}) for the Coulomb interaction
are given, respectively, by:
\begin{equation}
\epsilon_k=\frac{e^2}{\epsilon\ }\int^{\infty}_{-\infty}dy'\
\rho(y')\int^{\infty}_{-\infty}dy\ \frac{e^{-(y-k+y')^2}}{\sqrt{\pi}}\
\ln\ y^2 \
\end{equation}
and
\begin{equation}
V(q,\Delta k)=\frac{e^2}{\epsilon\ }\ \frac{1}{L}\
\frac{e^{-q^2/2}}{\sqrt{2\pi}}\int^{\infty}_{-\infty}dy\ e^{-(y-\Delta
k)^2/2}\ K_0(qy) \label{fullpot}
\end{equation}
where $e$ is the electron charge, $\epsilon$ is the dielectric constant, $q$ is
the
momentum transfer ($\delta$ in Eq. (\ref{h1d})),
$\Delta k=k-k'$, $\rho$ is the positive background density, and $K_0$ is a
modified
Bessel function. Notice that $e^2/\epsilon l_B$ (or $e^2/\epsilon$,
as we use units of $l_B=1$) is the natural energy scale in the problem.
In particular, the Hartree-Fock effective two body potential
between two particles with momenta $k_1$ and $k_2$ is given by
\begin{equation}
V_{HF}(k_1,k_2)=V_{H}(|k_1-k_2|)+V_{ex}(|k_1-k_2|)
\end{equation}
where the Hartree and exchange terms are obtained from Eq. (\ref{fullpot})
by setting $q \rightarrow 0$, $\Delta k=k_1-k_2$ and $q=k_1-k_2$, $\Delta k=0$
respectively (with a factor of $-1$ for the exchange):
\begin{eqnarray}
V_{H}(|k_1-k_2|)&=&-\frac{1}{2L}\frac{e^2}{\epsilon\ }\
\int^{\infty}_{-\infty}dy\ \frac{e^{-(y-k_1+k_2)^2/2}}{\sqrt{2\pi}}\
\ln y^2 \\ V_{ex}(|k_1-k_2|)&=&-\frac{1}{2L}\frac{e^2}{\epsilon\ }\
e^{-(\frac{k_1-k_2}{2})^2}\ K_0\left( (\frac{k_1-k_2}{2})^2 \right)\
.\nonumber
\end{eqnarray}
We have subtract a logarithmic divergence from the Hartree term
($\sim\ln q|_{q\rightarrow0}$), that is independent of $k_1$ and $k_2$,
and thus simply contributes to a constant in the energy.
The Hartree contribution to the effective two body potential, $V_H$, is
repulsive and long-ranged, whereas the one from exchange, $V_{ex}$, is
attractive and short-ranged. We will show that it is a balance between these
two effective interactions that controls the short length scale
behavior of the density distribution.

We will proceed by first presenting exact numerical results for a
small system, and then using these results to justify a picture that
the short length scale behavior of the density distribution is
controlled by the Hartree-Fock terms.

We study the edge structure of a system that we divide into ``edge''
and ``bulk'' electrons (See Fig. 4).  We consider just one edge, say,
the $R$ edge, and assume the ``bulk'' extends to infinity in the
opposite direction.  The occupation of the ``bulk'' levels is fixed to
be 1. Doing so, we can concentrate all the computations on the
``edge'', as the effect of ``bulk'' electrons is simply reduced to a
contribution to the one particle dispersion of the ``edge'' electrons.
Such division presents no harm, as long as the edge excitations under
consideration do not change the ``bulk'' occupation.

Consider a strip geometry with $L=20\ l_B$ and periodic boundary
conditions (cylinder). The ``edge'' is composed of 10 electrons in 20
single particle states. We start with a sharp background, and smoothen
it by changing the width $w$ in which the density drops from the bulk
value ($\rho=1/2\pi$, for $\nu=1$) to zero (see Fig. 5). The ``bulk''
electrons contribute to an additional term in the dispersion.  The
effect of adding a variation in the positive charge density over a
length scale $w$ can be thought of as simply superimposing a dipole to
the effective edge potential for a sharp edge, as shown in Fig. 5.

Fig. 6 displays the energy levels for different total momentum $K$ of
the edge electrons (the sites, numbered from 1 to 20, are assigned
$k=0$ to 19). For $w<8\ l_B$ the ground state has $K=45$, {\it i.e.},
all the electrons are packed up to one side and the edge is described
by the $Z_F=1$ Fermi liquid. For $w=9\ l_B$ the ground state is no
longer the sharp configuration with minimum $K=45$, but has moved to a
configuration with $K=60$ (see Fig. 7).  The occupation number
distribution is shown in Fig. 8 for $w=9$ and $10\ l_B$.  Notice the
formation of a lump of electrons distant $\sim 2\ l_B$ from the bulk
(for $L=20$, $\Delta k=1$ corresponds to a distance $2\pi/L\
\sim0.314\ l_B$).

The exact calculation for a small system seems to indicate that the
sharp edge is robust against the smoothening of the background charge,
up to a point where there is a transition, and the density
redistributes.  We will show that the robustness of the sharp edge is
a consequence of the attractive exchange, that tents to keep the edge
packed. Eventually, as the strength of the dipole resulting from the
smoothening of the background is increased, the short range attraction
due to exchange can no longer sustain the edge sharp, and a lump of
the electrons splits and forms a ``puddle'' near the minimum of the
effective potential seen by the QH Liquid. Let's illustrate the point
above by calculating the effective single particle energy within the
Hartree-Fock approximation for the distribution that has occupied
levels for all negative momenta, {\it i.e.}, a sharp $R$ edge (we have
centered the coordinate system on the edge).
\begin{equation}
\epsilon(k)=\epsilon_k+\Sigma_H(k)+\Sigma_{ex}(k)\ ,
\end{equation}
where
\begin{eqnarray}
\epsilon_k&=&\frac{e^2}{\epsilon\ }\int^{\infty}_{-\infty}dy'\
\rho(y')\int^{\infty}_{-\infty}dy\
\frac{e^{-(y-k+y')^2}}{\sqrt{\pi}}\ \ln\ y^2 \nonumber\\
\Sigma_H(k)&=&-\frac{e^2}{\epsilon\ }
\int^{0}_{-\infty}\frac{dk'}{2\pi}\int^{\infty}_{-\infty}dy\
\frac{e^{-(y-k+k')^2/2}}{\sqrt{2\pi}}\ \ln\ y^2 \nonumber\\
\Sigma_{ex}(k)&=&-\frac{e^2}{\epsilon\ }\int^{0}_{-\infty}\frac{dk'}{2\pi}\
e^{-(\frac{k-k'}{2})^2}\ K_0\left( (\frac{k-k'}{2})^2 \right)\ .\nonumber
\end{eqnarray}
One should notice that the electronic occupation number $n_k=\langle
c^\dagger_kc_k\rangle$ differs from the electronic charge density
$n(y)=\langle c^\dagger(y)c(y)\rangle$. The later is obtained from the
former by a convolution with a Gaussian of variance $\sigma^2=1/2$
(the single level charge distribution). With this in mind, one can
show that the background density that cancels the electronic charge
density is exactly the one that makes $\epsilon_k+\Sigma_H(k)=0$, as
it should be expect. A sharp background charge distribution as shown
in Fig. 5 makes a sharp electronic occupation distribution even more
stable, as we have an extra dipole term, resulting from the difference
between a sharp positive charge background and the electronic charge
density of a sharp occupation distribution, and which favors the
levels with negative $k$ to remain occupied.

As we change the background configuration, we alter the one particle
dispersion $\epsilon_k$.  For the stability of this edge it is
necessary that the effective single particle energy of any unoccupied
level be higher than the one of any occupied level. Fig. 9 shows the
effective single particle energy for unoccupied levels for different
values of the parameter $w$, which measures the width which it takes
the background to decrease from its bulk value to zero. The potential
obtained for $w=0\ l_B$ is primarily due to the exchange term, which
stabilizes a sharp edge. The exchange potential is short ranged,
reaching zero within $\sim 1.5 - 2\ l_B$; the overshoot for $w=0\ l_B$
is due to the dipole which results from the difference between a sharp
positive charge background and the electronic charge density of a
sharp occupation distribution, as mentioned previously, whose
contribution decays to zero as $1/|k|$ for large $|k|$. For $w \sim
11\ l_B$ the condition for stability is violated (the higher value for
the $w$ that marks the transition, as compared to the small system
result, can be regarded as due to a finite size effect, to the
Hartree-Fock approximation, or to both). It is then more advantageous
to move electrons to the minimum $\epsilon(k)$ locations. A simple
picture is that particles start to escape from the sharp edge and
start to deposit at a distance of order $\sim 2\ l_B$ away from the
initial boundary.  This separated lump brings in two new boundaries
into the problem. These new boundaries break our previous chiral
geometry, as we now have three Fermi points finitely separated. The
three Fermi points describe two right-moving branches and one
left-moving branch of edge excitations.  From this point on one should
expect that the interactions will destroy the Fermi Liquid
singularities, and we will have three Luttinger singularities.

Notice that we said the necessary condition for the stability of the
sharp edge is that $\epsilon(k)$ be larger for unoccupied states than
for occupied states.  But we have not yet argued it is sufficient. It
is possible that even if this condition is satisfied one can have a
ground state for the interacting problem different than the sharp
edge, as the energy could be lowered by rearranging many particles.
Worse, it is possible that the hopping terms, which couple different
states in configuration space (and are not included within
Hartree-Fock), would completely modify the picture.  In particular,
the Hartree-Fock approximation does not allow the distributions in
Fig. 1b and Fig. 1c. Therefore, we cannot use the Hartree-Fock
calculation alone to judge which of the distributions in Fig. 1 and
Fig. 2 is realized after the transition.  However, the exact
diagonalization results for small systems that we have presented
support the picture described in Fig. 2. This suggests the transition
is mainly controlled by Hartree-Fock terms, and that the conclusions
depicted from the single particle potential obtained within
Hartree-Fock seem to be qualitatively correct. Certainly we cannot
rule out the possibility that Fig. 1b and Fig. 1c might be realized
for some other interaction and edge potential.

To finalize this section, we present in Fig. 10 the spectrum and
occupation numbers calculated for the small system with $w=10\ l_B$,
but now within the Hartree-Fock approximation (the hopping or
off-diagonal elements were suppressed). Compare the spectrum to the
exact diagonalization for $w=10 \l_B$ displayed in Fig. 7. The
occupation numbers in the Hartree-Fock approximation suggest the
separation of part of the density from the main fluid, and the
appearance of two more singularities. The hopping elements would take
charge in redistributing the density, modifying the form of the
singularity.

\section{The Edge reconstruction for Quantum Dots}

The effect we describe in this paper is not particular to large
systems.  In fact, the exact results for small systems, which we used
to support the Hartree-Fock picture, directly indicate that the
transition occurs for finite systems. The edge density redistribution
is a consequence of the balance between the confining potential, the
repulsive Hartree term, and the short-ranged exchange, all of these
present regardless of the size of the system.

Quantum Dots are innately interesting systems for observing this
transition.  To begin with, because the number of electrons is small,
the redistribution will involve a substantial part of the total number
of particles in the dot, which can then be considered not simply an
edge effect, but, in a way, a bulk effect as well. Secondly,
experiments on resonant tunneling into Quantum Dots in the FQH regime
should be sensitive to a transition involving a redistribution of the
particle density, both because the energies of adding one electron to
the dot on both sides of the transition should differ (which can be
measured by the position of the resonant peaks), and because a change
in the size of the dot will change the coupling to the probe leads as
well. Third, the transition can be driven by altering the confining
potential, either changing the voltage on a back-gate, or changing the
magnetic field (which varies the radii of the orbits, and consequently
the potential seen by each orbit).

In this section we study some of the consequences of the transition as
applied to Quantum Dots. We study systems with total number of
particles up to $N_p= 70$. We will use only the Hartree-Fock matrix
elements, because both the system is not small enough for exact
diagonalization, and the Hartree-Fock approximation seemed to contain
the essential elements to describe the transition.

The energy eigenstates of the Hamiltonian in Eq. (\ref{h1d}) are also
eigenstates of total momentum. The true ground state is a
superposition of different occupation number states in Fock space, all
of them with the same total momentum. We will call hopping elements
the terms in the Hamiltonian that couple different states in Fock
space. The Hartree and Fock terms couple a configuration in Fock space
to itself. Within Hartree-Fock, {\it i.e.}, neglecting the hopping
terms in the Hamiltonian, any occupation number state is an energy
eigenstate.  Finding the ground state is then equivalent to
determining the configuration of particles that minimizes a classical
energy function. Notice that within Hartree-Fock all $\langle
c^{\dagger}_{\lambda} c_\lambda \rangle$ are equal to either 0 or 1.

We focus on a disk geometry, which is more appropriate for describing
a dot. The single particle states are labelled by the angular momentum
quantum number. The matrix elements are given by
\begin{equation}
V_{m1,m2,m3,m4}=\langle m1\ m3|\hat{V}|m2\ m4\rangle \ ,
\end{equation}
where the state $|m\ m'\rangle$ stands for a particle in the level
labelled by $m$, and another in the level $m'$ (\ $\langle z_1,z_2|m\
m'\rangle=
\phi_{m,m'}(z_1,z_2)=\frac{1}{\sqrt{\pi}}
\frac{z_1^m z_2^{m'}}{ \sqrt{m!\ m'!}}
e^{-\frac{|z_1|^2+|z_2|^2}{2}}$, with
$z_{1,2}=\frac{x_{1,2}+iy_{1,2}}{\sqrt{2}}$).  The matrix elements in
the classical energy function that couple states $m$ and $m'$ are
obtained from the Hartree and Fock terms:
\begin{equation}
V^{HF}_{m,m'}=V_{m,m,m',m'}-V_{m',m,m,m'}\ .
\end{equation}
To obtain these coefficients it is easier to work in a basis in which
$\hat{V}$ is diagonal. The $|l,n\rangle$ basis, in which $\langle
z_1,z_2|l,n\rangle=
\frac{1}{\sqrt{\pi}}\frac{z_+^n z_-^l}{ \sqrt{n!\ l!}}
e^{-\frac{|z_+|^2+|z_-|^2}{2}}$, where $z_{\pm}=\frac{z_1\pm
z_2}{\sqrt{2}}$, is such that $\langle l,n|\hat{V}|l',n'\rangle=V(l)\
\delta_{l,l'}\ \delta_{n,n'}$.  For the Coulomb interaction we have
\begin{eqnarray}
V(l)&=&\frac{1}{2}\frac{e^2}{\epsilon\ l_B}\int \frac{dz_+^2
dz_-^2}{\pi^2}\ \frac{1}{2|z_-|}
\frac{|z_+|^{2n} |z_-|^{2l}}{n!\ l!} e^{-|z_+|^2}e^{-|z_-|^2}\\
&=&\frac{e^2}{\epsilon\ l_B}\
\frac{1}{4}\frac{\Gamma(l+1/2)}{\Gamma(l+1)} \ .\nonumber
\end{eqnarray}
The Hartree and exchange terms are obtained, respectively, using
\begin{equation}
\langle m\ m'|\hat{V}|m\ m'\rangle=\sum_{l,n}\
V(l)\ |\langle m\ m'|l,n\rangle|^2
\end{equation}
and
\begin{equation}
\langle m\ m'|\hat{V}|m'\ m\rangle=\sum_{l,n}\
(-1)^l\ V(l)\ |\langle m\ m'|l,n\rangle|^2\ .
\end{equation}
The confining potential is assumed to be parabolic,
$\epsilon(r)=\frac{1}{2}kr^2$, with $r$ the distance from the center
of the dot, and $k$ the strength of the confining potential. It is
convenient to write $k=\alpha_0\frac{e^2}{\epsilon l_B}l^{-2}_B$, so
that $\alpha_0$ is a dimensionless parameter, maintaining
$e^2/\epsilon l_B$ and $l_B$ as our, respectively, energy and length
units. The single particle dispersion is given by
\begin{eqnarray}
\epsilon_m&=&\frac{1}{2}\ \alpha_0 \ \frac{e^2}{\epsilon\ l_B}\
\int \frac{dz^2}{\pi}
\ 2|z|^2\frac{|z|^{2m}}{m!}\  e^{-|z|^2}\\
&=&\alpha_0 \ \frac{e^2}{\epsilon\ l_B}\
\frac{\Gamma(m+2)}{\Gamma(m+1)}= \alpha_0\ \frac{e^2}{\epsilon\ l_B}\
(m+1)\ .\nonumber
\end{eqnarray}
The energy function that must be minimized is
\begin{equation}
{\cal E}=\sum_{m}\epsilon_m\ n_m\ +\ \sum_{m,m'}V^{HF}_{m,m'}\ n_m\
n_m' \ ,
\end{equation}
where the $n_m$'s are 0 or 1, constrained to $\sum n_m=N_{p}$, the
total number of particles. We obtained numerically $V^{HF}_{m,m'}$ for
the first 80 levels ($0\le m,m'\le 79$), and searched for the minimum
of ${\cal E}$ for different values of $N_p$ and $\alpha_0$. We find
that, depending on these parameters, the minimum energy configuration
switches from a compacted to a separated droplet.

In Fig. 11 we display the occupation of the orbits as function of
$\alpha_0$ for $N_p=60$ (occupied orbits are displayed in black, and
unoccupied ones in white).  For strong confining potentials the
occupied levels are the ones with minimum angular momentum. As the
confining strength is decreased, there is a transition, and unoccupied
levels inside the dot appear. After the transition, hopping elements
become important, and take charge in redistributing the occupation,
which can then have partially filled levels. Fig. 12 displays the
orbital occupation for fixed $\alpha_0$, with $N_p$ varying from 70 to
30 electrons, where there is also a transition, with a separated
droplet for smaller systems.

We would like to point out that the exchange term is of key importance
in order to have a compacted dot solution. Notice that we have assumed
that occupation is non-zero only for the first Landau level, and that
the spins are fully polarized. One could argue that these assumptions
alone can lead to a compacted drop, as a strong enough confining
potential can always be chosen such that the ground state is the
minimum total angular momentum solution even if one take only the
repulsive Hartree term. This would be possible because the particles
would be squeezed to the center, without being able to occupy higher
Landau levels, or flip spin (see Fig. 13, where we repeat the
calculation for Fig. 11 without the exchange term).  In reality, as we
increase the confining potential, there are two mechanisms which tend
to lower the total energy of the dot, one by compacting the particles
to the low angular momentum orbits in the first Landau level, with
polarized spins, and the other by moving particles to a higher Landau
level or opposite spin polarization state. By increasing the confining
potential we enhance both of these effects.  Therefore, in order to
have a $\nu=1$ compacted dot we must have a window of $\alpha_0$ that
allows totally packed dots with no higher Landau level occupation.  It
is here that exchange comes in play, providing an attractive
interaction that lowers the bound on $\alpha_0$ to have a compacted
dot, which opens that window.

The ideas above can be expressed quantitatively. The lower bound on
$\alpha_0$, {\it i.e.}, the minimum confining strength necessary to
keep an $N_p$-particle dot compacted, can be obtained as follows.
Within the Hartree approximation, $\alpha^{min}_0$ is obtained from
the condition that the net electric field on the edge of the dot due
to the electrons just balances the field due to the confining
potential. The radial field due to the electrons diverges
logarithmicaly, $E_r\propto
\frac{e}{\epsilon l^2_B}\ln (R/\lambda_c)$, where $R$ is the radius of the
droplet, and $\lambda_c$ is an ultraviolet cut-off length scale. The
field due to the confining potential is $E_r=-\frac{e}{\epsilon
l^2_B}\alpha_0 R/l_B$, so we find that $\alpha^{min}_0\propto
N_p^{-1/2} \ln (\frac{N_p}{\lambda_c^2/2 l_B^2})$. Indeed, this
dependence of $\alpha_0$ on $N_p$ fits very well the numerical results
obtained when the exchange term is omitted (Hartree approximation),
where we find $\alpha^{min}_0\sim 0.118 N_p^{-1/2} \ln
(\frac{N_p}{0.16})$ (see Fig.  14a). We find that a similar function
dependance on $N_p$ reasonably fits the results obtained within
Hartree-Fock, with $\alpha^{min}_0\sim 0.083 N_p^{-1/2} \ln
(\frac{N_p}{0.21})$ for our range of $N_p$ (Fig. 14b). Notice that the
attractive exchange term has the tendency to keep the dot compated,
lowering the value of $\alpha^{min}_0$.

The upper bound on $\alpha_0$ can be obtained by estimating the energy
decrease of moving one particle from the edge to the center. The
electrostatic energy of a disk of radius $R$ and density
$\rho_0=\frac{e}{2\pi\ l_B^2}$ is ${\cal E}_{disk}=\frac{e^2}{\epsilon
l_B}
\frac{2}{3\pi}(R/l_B)^3$, which gives an estimate for the electrostatic
energy of adding one electron to the edge of the dot of $\delta{\cal
E}_{edge}=
\frac{e^2}{\epsilon l_B}\frac{2\sqrt{2}}{\pi}\sqrt{N_p}$.
The electrostatic energy cost of adding an electron to the center of
the disk is $\delta{\cal E}_{center}=\frac{e^2}{\epsilon
l_B}\sqrt{2}\sqrt{N_p}$. The total decrease in energy of moving one
particle from the edge to the center is $\Delta{\cal
E}=\frac{e^2}{\epsilon l_B}\left(\alpha_0 N_p -
\frac{\pi-2}{\pi}\sqrt{2}\sqrt{N_p}\right)$, which has to be $<\hbar
\omega_c$ (or $<{\cal E}_{Zeeman}$) if we want to have occupation
solely in the first Landau level (or with polarized spins). So we find
$\alpha^{max}_0\sim\ \frac{\pi-2}{\pi}\sqrt{2} N_p^{-1/2} +
\frac{\hbar\omega_c}{e^2/\epsilon l_B}({\rm or}\ \frac{{\cal
E}_{Zeeman}}{e^2/\epsilon l_B})\ N_p^{-1}$.  The last term contains
the ratio between the cyclotron (or Zeeman) and Coulomb energies. The
condition for a compacted dot is
$\alpha^{min}_0<\alpha_0<\alpha^{max}_0$.  For GaAs
$\frac{\hbar\omega_c}{e^2/\epsilon l_B}\sim0.4\sqrt{B}$, with $B$ the
magnetic field in Tesla. So for reasonable values of $B$ and $N_p$ in
a dot, the term $N_p^{-1/2}$ in $\alpha^{max}_0$ is the dominant one,
which constrains the maximum possible $N_p$ to the one that makes
$\alpha^{min}_0\sim\alpha^{max}_0$, which gives $N_p^{max}\sim 106$.
If we use the value for $\alpha^{min}_0$ given by the Hartree term
alone we find $N_p^{max}\sim 12$. Although these values are rough
estimates, they should make it clear that the attractive exchange term
plays a major role in opening up a window in $\alpha_0$ for which
there is a compacted dot solution.

We have also performed the Hartree-Fock calculation for spin 1/2
electrons with Zeeman energy ${\cal E}_{Zeeman}\rightarrow 0_+$
(simply to break the degeneracy between the two spin polarized
configurations).  For large confining potentials, both spin up and
down electrons form compact droplets of filling fraction $\nu=1$.
However the droplet of, say, spin down electrons is smaller, and the
electrons near the edge form a ferromagnetic state as pointed out in
Ref. \cite{Hal}.  As we decrease $\alpha_0$, the separation between
the spin-up edge and the spin-down edge increases.  At even smaller
$\alpha_0$ the spin-down electrons non-longer form a compact droplet
which maybe a sign of FQH states. As $\alpha_0$ decreases below
$\alpha_s\sim \ 0.53\ N_p^{-1/2} + 0.49\ N_p^{-1}$, all the electrons
are spin polarized (the coefficient in $N_p^{-1/2}$ is approximatly
the same obtained from the electrostatic consideration, and the one in
$N_p^{-1}$ shows the tendency of exchange to align spins as an
effective Zeeman energy).  Thus for
$\alpha^{min}_0<\alpha_0<\alpha_s$, the electrons form a spin
polarized compact droplet. The exchange term tends to align the spins,
increasing further the size of the window of $\alpha_0$'s such that
the drop is compact.

The next question is how to experimentally obtain a value for
$\alpha_0$ that falls within the window above. In order to make the
connection to real samples, we use the parabolic confining potential
in Ref. \cite{Wingreen}.  There they use
$V_{ext}(r)=\frac{1}{2}m^*\omega^2r^2$, with $m^*$ the effective
electron mass in GaAs, and $\hbar\omega=1.6{\rm meV}$ for the
particular device.  The relation between our dimensionless $\alpha_0$
(or alternatively, given in units of $\frac{e^2}{\epsilon l_B}=1$ and
$l_B=1$) to this confining potential is obtained by equating
$\alpha_0\frac{e^2}{\epsilon l_B}l^{-2}_B$ to $m^*\omega^2$, which
gives $\alpha_0=\frac{(\hbar\omega)^2}{(e^2/\epsilon
l_B)(\hbar\omega_c)}$, or $\alpha_0\sim 0.376\ B^{-3/2}$, $B$ in
Tesla. For $N_p=40$, for example, we find that $B=2.5$T will yield a
value of $\alpha_0$ in the allowed window, so that the dot occupation
will be the one of a compacted $\nu=1$ droplet. As we increase $B$
beyond 3.1T the edge will undergo a reconstruction.

We now turn into the possibility of probing experimentally the
transition between a compacted and an expanded dot. In order to make a
clear connection between this expansion effect and experimentally
observable quantities, we describe below the implications of the
effect to tunneling experiments into Quantum Dots. In resonant
tunneling experiments, the energy difference between the ground states
of an $N+1$ and $N$ electron system, $\mu(N)$, can be probed by
tunneling in and out of the dot a single electron at a time, when the
Fermi level of the electrodes become resonant with the quantum level
of the dot \cite{Su,McEuen,Ashoori}.  By following a peak, the
dependence of the chemical potential on the magnetic field can be
observed.  In Fig. 15 we calculated, within the Hartree-Fock
approximation, this dependence of the chemical potential (in meV) on
$B$ (in T) for dots with $N_p$ from 35 to 38 . The ``sawtooth'' for
$B<2.5$T corresponds to spin down electrons being flipped and moved
from the center to the edge of the dot. The region between roughly
2.5T and 3T is the window of magnetic fields for which the electrons
form a compact $\nu=1$ droplet. The ``dislocation'' near $B=3$T
corresponds to the reconstruction of the electron number occupation,
marking the transition from the compacted to the expanded
configuration.

In addition to this anomaly in the peak position vs. $B$, the
expansion of the size of the dot will also increase its coupling to
the probing leads, as this coupling depends on the distance between
leads and island. The tunneling current should then increase for an
expanded dot. In Fig. 16 we show the dependence of the size of the
droplet (measured as the radius of the orbit of the outmost electron)
on $\alpha_0$ and on the number of particles $N_p$ in the dot. In Fig.
16a we show the size of a 60 electron droplet as a function of
$\alpha_0$, and in Fig. 16b we fixed the value of $\alpha_0$ and
varied the number of electrons from 40 to 70. The effect on the
amplitute of the resonant peak, together with the anomaly in the peak
position, should be a signature that a transition is indeed occurring
in the occupation density of the Quantum Dot.

\section{The Edge Modes After the Transition}

As we have seen in the previous sections, after smoothening the edge
potential enough, a transition takes place, and the Fermi Liquid
occupation density gives way to a more complex state. In terms of the
electronic occupation distribution, the new state looks as if
electrons start to deposit a certain distance away from the bulk of
the QH Liquid. The QH ``puddle'' that is formed, as mentioned above,
brings in two more boundaries for each edge, and we then have three
singularities. The interactions take charge in destroying the Fermi
Liquid discontinuity, as the three singularities are finitely
separated.

These three singularities can be related to three branches of gapless
modes, which correspond to particle-hole excitations near each of the
singularities. An intuitive way to visualize the three branches is by
considering the occupation number after the transition simply within
Hartree-Fock, which would look as in Fig. 17. There, the three edges
are clearly identified. Correlations destroy the Fermi
discontinuities, but we will still have the three Luttinger Liquid
singularities at the three Fermi points.  Notice that the pair of
branches that is added always have opposite chirality. An edge that
had one right-moving branch before the transition, for example, will
have two right-moving branches and one left-moving branch. All these
three branches are strongly coupled.  One clear experimental
consequence of now having one branch moving in the opposite direction
(the one left-moving branch in the originally right-moving edge, for
example) is that one could probe such excitations, where originally
there was none. We will show, however, that the presence of impurities
localize two of the three branches, and such back-propagating modes
cannot be observed beyond the localization length.

We will use the bosonized description of the edge states in the FQH
regime presented in Ref. \cite{XGW}, and which we summarize below for
our particular case of $\nu=1$. Let $\phi_{R,L}$ be two fields,
described by the Lagrangian density
\begin{equation}
{\cal L}_{R,L}=\frac{1}{4\pi}\ \partial_x\phi_{R,L}\ (\pm \partial_t-
v \partial_x)\phi_{R,L}
\end{equation}
($v$ is the velocity of the excitations) and the equal-time commutation
relations
\begin{equation}
[\phi_{R,L}(t,x)\ ,\ \phi_{R,L}(t,y)]=\pm i\pi \ sgn(x-y)\ .
\end{equation}
Left and right moving electron operators can be written as
$\Psi_{R,L}(x,t)=:e^{\pm i\phi_{R,L}(x,t)}:$, which can be shown to
satisfy the correct anticommutation relations \cite{Jackiw}.  The
electron density is given by $\rho_{R,L}=\partial_x\phi_{R,L}$, and
the Hamiltonian is
\begin{equation}
H_{R,L}=\frac{v}{4\pi}\int dx\  \rho^2_{R,L}\ .
\end{equation}

Consider three edge branches as depicted in Fig. 18, labelled by
$i=1,2,3$.  We can generalize the description above to include several
branches, writing the following Lagrangian density:
\begin{equation}
{\cal L}=\frac{1}{4\pi}\ \sum_{i,j}\left[ K_{ij}\
\partial_t\phi_i\partial_x\phi_j- V_{ij}\
\partial_x\phi_i\partial_x\phi_j\right]\ ,\label{lagrangian}
\end{equation}
where the matrices $K$ and $V$ contain, respectively, information on
the direction of propagation (chirality) of each branch, and
interactions between the branches (including a diagonal term
containing the velocities).  For this analysis let us assume two $R$
branches (1 and 3) and one $L$ branch (2), such that
\begin{equation}
K=\left(\matrix{1&0&0\cr0&-1&0\cr0&0&1}\right)\ .
\end{equation}
Electron operators can be written as
\begin{eqnarray}
\Psi_L&\propto& e^{i\sum_i l_i \phi_i}\\
l_i&=&\sum_j K_{ij}L_j \ , \nonumber  \\
\end{eqnarray}
where the $L_i$'s are integers satisfying $\sum_iL_i=1$.  The
Lagrangian in (\ref{lagrangian}) is not the most general one.  We left
out four fermion terms that cannot be written as the product of two
densities. We know that it is the interactions between the densities
that are responsible for changing the Fermi edge singularity, and here
we will concentrate on this particular effect of interactions.  We
have also assumed a local interaction in the densities. However, the
long range interaction can be easily included by allowing $V_{ij}$ to
have momentum dependence. In particular, the long range Coulomb
interaction contributes to a term $\sum_k \lambda_k (\sum_i
\rho_{k,i}) (\sum_i \rho_{-k,i})$, where $\lambda_k\propto \ln k$ is
the Fourier transformation of the $1/r$ Coulomb interaction and
$\rho_{k,i}$ the Fourier component of the density of the i$^{th}$
branch. We see at long distances the most important interaction term
is the one that involves only the total charge density $\sum \rho_i$.
This contribution is
\begin{equation}
\frac{\lambda}{4\pi}\int dx\ \left(\sum_i\rho_i\right)^2=
\frac{\lambda}{4\pi}\int dx\ \sum_{i,j}\rho_i \rho_j \ ,
\end{equation}
which, when summed to the velocity terms, gives the total $V$ matrix
\begin{equation}
V=\left(\matrix{v_1+\lambda&\lambda&\lambda\cr\lambda&v_2+
\lambda&\lambda\cr\lambda&\lambda&v_3+\lambda}\right)\
{}.
\end{equation}
Here, for simplicity, we have assumed that $\lambda$ is a large
constant independent of momentum. This will be the case if the Coulomb
interaction is screened at a long distance ({\it e.g.} by gates
nearby).  We will focus primarily in the case where $\lambda\gg v$'s,
{\it i.e.}, strongly coupled branches. Also, if the system has
particle-hole symmetry, then $v_1=v_3$.  We start by rewriting the
Lagrangian (\ref{lagrangian}) in terms of new fields
$\tilde{\phi}_i=\sum_jU_{ij}\ \phi_j$ that simultaneously diagonalize
$K$ and $V$.  Furthermore, we would like to keep, for convenience,
\begin{equation}
\tilde{K}=(U^{T})^{-1}KU^{-1}=K
\end{equation}
so that the commutation relations of the $\tilde{\phi}$'s are the same
as the ones for the $\phi$'s. The transformation matrix $U$ for
$\lambda\gg v$'s is
\begin{equation}
U_{\frac{v}{\lambda}\rightarrow 0}=\left(\matrix{1&1&1\cr
\frac{1}{\sqrt{2}}&\sqrt{2}&\frac{1}{\sqrt{2}}\cr \label{Umatrix}
-\frac{1}{\sqrt{2}}&0&\frac{1}{\sqrt{2}}}\right) \ .
\end{equation}
The mode in the first line of $U$,
$\tilde{\phi}_1=\phi_1+\phi_2+\phi_3$, is simply the total charge
density mode (equivalently, $\tilde{\rho}_1=\rho_1+\rho_2+\rho_3$, as
$\rho=\partial_x \phi$). What we show next is that, once we add to the
Lagrangian (\ref{lagrangian}) electron scattering terms due to the
presence of impurities, this total charge mode is left unperturbed,
and the other two will localize.

The scattering terms between electron operators in the three edges
that can be added to the Lagrangian are bosonic couplings with zero
charge. These can be written as $T_L=e^{i\sum_i l_i
\phi_i}=e^{i\sum_{ij} L_i K_{ij} \phi_j}$, where now the $L_i$'s are
integers satisfying $\sum_iL_i=0$ ($T_L$ is bosonic and neutral). In
terms of the rotated fields $\tilde{\phi}$'s, $T_L=e^{i\sum_i
\tilde{l}_i \tilde{\phi}_i}$, where $\tilde{l}_i=\sum_j l_j
U^{-1}_{ji}$.  Let us focus on the most relevant $T_L$'s; the naive RG
dimension can be obtained from the $T_L$ correlations
\begin{eqnarray}
\langle T^{\dagger}_L(t=0,x)\ T_L(t=0,x=0)\rangle&\propto&e^{-\sum_{i,j}
\tilde{l}_{i} \tilde{l}_{j} \ \langle \tilde{\phi}_i(t=0,x)\
\tilde{\phi}_j(t=0,x=0)\rangle} \\
&\propto& x^{-\sum_{i} \tilde{l}_{i}^{2}}=x^{-\gamma_L}\nonumber\ ,
\end{eqnarray}
Writing $\gamma_L$ in terms of the $L$'s we obtain:
\begin{eqnarray}
\gamma_L&=&\sum_{i}\tilde{l}_i^2=\tilde{l}^T\tilde{l} \label{GammaL} \\
&=&l^T U^{-1}(U^{-1})^T l \nonumber \\ &=&L^T K U^{-1}(U^{-1})^T K L
\nonumber \\ &=&L^T U^T (U^T)^{-1} K U^{-1} (U^{-1})^T K U^{-1} U L
\nonumber \\ &=&L^T U^T \tilde{K}^2 U L \nonumber \\ &=&L^T (U^T U) L
\ , \nonumber
\end{eqnarray}
where we used $\tilde{K}^2=K^2=I$. It is easy to show that the minimum
$\gamma_L$, with $\sum_iL_i=0$, are given by:
\begin{equation}
L=\pm \left(\matrix{1\cr -1\cr 0}\right)\ {\rm and}\ L=\pm
\left(\matrix{0\cr -1\cr 1}\right)\ ,
\end{equation}
which correspond to $T_L$ operators that transfer charges between the
center branch ($L$) to the two side branches ($R$). In terms of the
$\tilde{l}$'s, we have:
\begin{equation}
\tilde{l}=(U^{-1})^T K L=K U L=\pm \left(\matrix{0\cr \frac{1}{\sqrt{2}}\cr
-\frac{1}{\sqrt{2}}}\right)\  {\rm and}\
\pm \left(\matrix{0\cr \frac{1}{\sqrt{2}}\cr \frac{1}{\sqrt{2}}}\right)\ .
\end{equation}

The Hamiltonian density with these most relevant $T_L$ terms added is
\begin{equation}
{\cal H}={\cal H}_0+{\cal H}_T
\end{equation}
where
\begin{equation}
{\cal H}_0=\frac{\tilde{v}_1}{4\pi}\ (\partial_x\tilde{\phi}_1)^2 +
\frac{\tilde{v}_2}{4\pi}\ (\partial_x\tilde{\phi}_2)^2 +
\frac{\tilde{v}_3}{4\pi}\ (\partial_x\tilde{\phi}_3)^2
\end{equation}
and
\begin{equation}
{\cal H}_T=
\xi_{+}(x) e^{i\frac{\tilde{\phi}_2+\tilde{\phi}_3}{\sqrt{2}}} +
\xi_{-}(x) e^{i\frac{\tilde{\phi}_2-\tilde{\phi}_3}{\sqrt{2}}} + H.c.
\end{equation}
The $\xi_{\pm}$ describe the random tunneling coupling due to
impurities, with correlations
$\langle\xi_{\pm}(x)\xi_{\pm}(y)\rangle=\Delta_{\pm}\delta(x-y)$.

Notice that $\tilde{\phi}_1$ remains free, as the added tunneling
terms do not depend on it. This one component (total charge, as
$\tilde{\rho}_1=\rho_1+\rho_2+\rho_3$) behaves just like the one
branch before the transition. The other two, we will argue below,
should be localized because of the impurities.

Let $\tilde{\phi}_{\pm}=(\tilde{\phi}_2\pm\tilde{\phi}_3)/\sqrt{2}$,
which obey the commutation relations
\begin{eqnarray}
\left[\tilde{\phi}_{\pm}(t,x)\ ,\ \tilde{\phi}_{\pm}(t,y)\right]&=& 0 \\
\left[\tilde{\phi}_{\pm}(t,x)\ ,\ \tilde{\phi}_{\mp}(t,y)\right]&=&-i\pi\
sgn(x-y) \nonumber
\end{eqnarray}
We can identify $\tilde{\Pi}_{\pm}=\frac{\partial_x
\tilde{\phi}_{\mp}}{2\pi}$ as the conjugate momenta to
$\tilde{\phi}_{\pm}$.  We can rewrite the Hamiltonian for
$\tilde{\phi}_2$ and $\tilde{\phi}_3$ in terms of
$\tilde{\phi}_{\pm}$:
\begin{equation}
{\cal H}_{2,3}=
\frac{\tilde{v}}{4\pi}\ \left[(\partial_x\tilde{\phi}_+)^2 +
(\partial_x\tilde{\phi}_-)^2\right] +
\xi_{+}(x) e^{i\tilde{\phi}_+} +
\xi_{-}(x) e^{i\tilde{\phi}_-} + H.c.
\end{equation}
where we assume that $\tilde{v}_2\approx\tilde{v}_3\approx\tilde{v}$.
This is a more complicated version of a Sine-Gordon (SG) Hamiltonian
density with position dependent coupling, as it involves
self-interactions in both a field and its conjugate momentum.

If we had only one of $\xi_+$ or $\xi_-$, we would have a Hamiltonian
for a simple SG with position dependent coupling, which we could write
as
\begin{equation}
{\cal H}=\frac{\tilde{v}}{4\pi}\left[ (2\pi\tilde{\Pi})^2 +
(\partial_x\tilde{\phi})^2 \right] +
\xi(x) e^{i\tilde{\phi}} + H.c.\ .
\end{equation}
Working in units of $\tilde{v}=1$, and rescaling the fields as
$\Pi'=\sqrt{2\pi}\tilde{\Pi}$ and $\phi'=\tilde{\phi}/\sqrt{2\pi}$
(which keep the commutation relations unchanged), we have
\begin{equation}
{\cal H}=\frac{1}{2}\left[ \Pi'^2 + (\partial_x\phi')^2 \right] +
\xi(x) e^{ig\phi'} + H.c.\ ,
\end{equation}
with $g=\sqrt{2\pi}$.  This problem, equivalent to a Coulomb gas with
position dependent chemical potential, was studied in refs.
\cite{Apel,Shulz}. The impurity coupling $\Delta_\pm$ is relevant for
$g<\sqrt{6\pi}$ (in the constant coupling constant or chemical
potential Coulomb gas, the condition is $g<\sqrt{8\pi}$). This is
indeed our case, and therefore the presence of impurities localizes
the other two branches of excitations represented by $\tilde{\phi}_2$
and $\tilde{\phi}_3$.

Notice that what we have done above is equivalent to understanding the
RG flows in the planes $\Delta_+=0$ and $\Delta_-=0$, and this implies
$\Delta_\pm$ are relevant in all directions around $\Delta_\pm=0$ if
$g<\sqrt{6\pi}$.  The RG flows to a strong fixed point when both
$\Delta_\pm \neq 0$.  It is possible that this strong fixed point is a
localized state, motivated by the flow when $\Delta_-=0$.  The
properties of this strong fixed point will be the subject of further
studies.

\section{Tunnelling into reconstructed edges}

The formalism in the last section can be used to study the electron
tunneling into the reconstructed edges. The electron propagator in
time in general has a form $$ <c^\dagger(t) c(0) >\propto {1\over
t^{\gamma_L}} $$ with $\gamma_L$ given in Eq. (\ref{GammaL}). But now
the $L_i$'s satisfy $\sum_i L_i=1$.  In the limit $\lambda/v \gg 1$
$U$ is given by Eq. (\ref{Umatrix}).  The minimum exponent is
$\gamma_L=1$ (the Fermi liquid value) for electrons described by
$L=(1,-1,1)$. This configuration corresponds to adding two electrons
on the two side branches and removing one electron from the center
branch.  Adding a single electron to the side branch leads to a
exponent $\gamma_L=2$ and to the center branch $\gamma_L=3$.

Let us consider tunneling between two reconstructed edges. At very low
temperatures and low voltages, the electron with the configuration
$L=(1,-1,1)$ will dominate the tunneling and leads to a linear $I-V$
curve \cite{tunpaper}, since $I\propto V^{2\gamma_L -1}$ and
$\gamma_L=1$.  At higher voltages, depending on the sample geometry,
it may be easier for an electron to just tunnel into the side branch
(with configuration $L=(1,0,0)$). In this case $I\propto V^3$ and
$(dI/dV)_{V=0}\propto T^2$.

The above discussion also apply the reconstructed edges of Laughlin
states of filling fraction $1/m$. But now for the $L=(1,-1,1)$
electron the exponent $\gamma_L=m$. $\gamma_L=2m$ for $L=(1,0,0)$ and
$\gamma_L=3m$ for $L=(0,1,0)$. We see that in the limit $\lambda/v \gg
1$ the minimum exponent in the electron propagator is not affected by
the edge reconstruction. This result is valid even when more then one
pair of edge branches are generated.  This is because adding electrons
of configuration $L=(1,-1,1,-1,...,1)$ just displace all the edge
branches by the same amount. Thus the electron of
$L=(1,-1,1,-1,...,1)$ just couples to the total density $\sum \rho_i$
and do not couple to other neutral modes.  In the limit $\lambda/v \gg
1$, the total density mode decouples from other neutral modes. This is
the reason why the $L=(1,-1,1,-1,...,1)$ electron always have the
exponent $\gamma_L=m$.  We would like to stress that the above result
is valid only at low energies (energies below the smallest Fermi
energy of generated edge branches). The high energy behavior of the
electron propagator is not clear. In that case it is probably better
to view the edge region as a compressible gas.

For tunneling between two reconstructed FQH edges, we expect
$I\propto V^{2m-1}$ and $(dI/dV)_{V=0}\propto T^{2m}$ at low
voltages and low temperatures. This is consistent with a recent
experiment on tunneling between (smooth) edges of $1/3$ FQH states.
\cite{IBMexp}

\section{Conclusion}

In this paper we studied the electronic density of Quantum Hall
Liquids, focusing on short length scales that are comparable with the
magnetic length. We found that sharp electronic occupation densities,
corresponding to a $Z_F=1$ Fermi Liquid, is possible because of the
chiral nature of the system. This sharp distribution is stable against
variations in the confining potential up to a certain point, beyond
which it undergoes a transition and electrons start to separate from
the bulk and deposit a distance $\sim 2\ l_B$ away. The transition is
shown to be qualitatively described within the Hartree-Fock
approximation. The separation generates a pair of branches of edge
states that move in opposite directions.

We would like to remark that the separated electrons do not form any
fractional quantum Hall state. This is because the separation between
the Fermi edges is always of order magnetic length for realistic
potentials.  In this case $1/3$, $1/5$,... states are all described by
Luttinger liquid and are indistinguishable.

We also present results for Quantum Dots, where we predict that this
effect of edge separation can be related to an expansion of the dot,
which could be experimentally observed.

The authors would like to thank Dmitrii Chklovskii, Yong Baek Kim,
David Abusch, Olivier Klein, Paul Belk, Marc Kastner, and Ray Ashoori
for useful discussions. This work is supported by the NSF Grant No.
DMR-91-14553.

\newpage

\begin{center}
{\large FIGURE CAPTIONS}
\end{center}
\
Figure 1. Possible singularities for the momentum occupation
distribution of a $\nu=1$ state: (a) $Z_F=1$ Fermi Liquid singularity,
(b) $Z_F<1$ Fermi Liquid singularity, and Luttinger Liquid
singularity.

\

Figure 2. Momentum occupation distribution after reconstruction, with
three singularities. The addition of two singularities correspond to
the addition of two branches of opposite moving edge excitations.

\

Figure 3. Cylindrical geometry, equivalent to a strip of length $L$
and periodic boundary conditions, where it is convenient to use the
Landau Gauge. The QH Liquid (shaded area) lies on the surface of the
cylinder, between its left and right edges at $\lambda_L$ and
$\lambda_R$.

\

Figure 4. The $\nu=1$ droplet is divide into ``bulk'', where all the
states are fully occupied, and ``edge'', in which states can be
partially occupied. The division allows one to focus the computations
solely on the ``edge'' electrons, with the ``bulk'' simply
contributing to the one particle dispersion. This sort of division is
not unique, as one can adjust the position of the boundary between the
two regions; this boundary can be moved as long as the sites on the
``edge'' side of it are all fully occupied.

\

Figure 5. Smoothened background charge density, which over a width $w$
drops from its bulk value to zero. Such density can be written as the
superposition of a sharp density profile tp a dipole term, which is
used to tune the confining potential as function of $w$.

\

Figure 6. Energy eigenstates obtained as function of the total
momentum $K$ of the edge electrons for values of $w$ ranging from 0 to
$7\ l_B$. Notice that the ground state for this range has $=K=45$, the
minimum momentum configuration for 10 electrons (the 20 sites used in
the calculation are assigned $k$ values from 0 to 19). Also, notice
that the energy levels of states of higher $K$ are pulled down as $w$
increases.

\

Figure 7. For $w\ge 8\ l_B$ the ground state configuration is no
longer the sharp edge distribution. The transition occurs near $w=8\
l_B$, and for $w= 9\ l_B$ is already fully developed, with the ground
state momentum moved to $K=60$.

\

Figure 8. Occupation numbers for the ground state past the transition,
for $w=9$ and $10\ l_B$. Notice how a lump of density moves away from
the main body of QH fluid.  This density profile (which goes from its
bulk value to zero by first decreasing, then increasing, and finally
decreasing again) is the signature of the existence of now three
singularities, and thus three branches of edge excitations.

\

Figure 9. The effective single particle potential calculated within
the Hartree-Fock approximation for different widths $w$, decreasing
from $0\ l_B$ to $15\ l_B$ in steps of $1\ l_B$. Notice that starting
at $w \sim 11\ l_B$, the condition for stability of a sharp edge is
violated, as there are locations with smaller effective potential than
the one at the edge of the sharp occupation density.

\

Figure 10. Energy levels and occupation numbers for $w=10 \l_B$
calculated within the Hartree-Fock approximation (the hoping or
off-diagonal elements were suppressed).

\

Figure 11. Occupation number of the angular momentum states as a
function of $\alpha_0$ for $N_p=60$, calculated within the
Hartree-Fock approximation.  The occupied orbits are represented in
black, and the unoccupied ones are shown in white.

\

Figure 12. Occupation number of the angular momentum states as a
function of $N_p$ for $\alpha_0=6.25\times 10^{-2}$, calculated within
the Hartree-Fock approximation.

\

Figure 13. Occupation number of the angular momentum states as a
function of $\alpha_0$ for $N_p=60$, calculated within the Hartree
approximation.

\

Figure 14. Minimum $\alpha_0$ necessary to keep an $N_p$-particle
droplet compacted, calculated using the Hartree (a) and Hartree-Fock
(b) approximation.  The solid line is the best curve fit consistent
with an electrostatic (Hartree) argument.

\

Figure 15. Dependence on the magnetic field $B$ of the energy cost to
add one more particle, $\mu$, to an island with 35 (lowest curve), 36,
37 and 38 (highest curve) electrons. The ``sawtooth'' corresponds to
$1<\nu<2$, where electrons are spin flipped and taken from the center
to the edge as the magnetic field is increased. The ``dislocations''
near $B\sim 3$T correspond to the transition between compacted and
separated dots. The region in between ($B\sim 2.5$ to $3$T) is the
window for which the dot is a compact $\nu=1$ droplet. These results
were obtained for the parabolic confining potential of the sample
studied in Ref. \cite{Wingreen}, with $\alpha_0=0.376 B^{-3/2}$ ($B$
in Tesla).

\

Figure 16. Radius of the QH droplet for (a) fixed $N_p$ and varying
$\alpha_0$, and (b) fixed $\alpha_0$ and varying $N_p$. One can vary
$\alpha_0$ by changing the strength of the confining potential, or by
changing the magnetic field.  For the parabolic confining potential of
the sample studied in Ref. \cite{Wingreen}, the parameter
$\alpha_0=0.376 B^{-3/2}$ ($B$ in Tesla).

\

Figure 17. Momentum occupation distribution after reconstruction, if
calculated only within the Hartree-Fock approximation. The addition of
two singularities correspond to the addition of two branches of
opposite moving edge excitations.

\

Figure 18. Three branches of edge excitations, two right-moving (1 and
3, on the sides) and one left-moving (2, in the center).

\end{document}